\documentstyle[11pt]{article}

\newtheorem{Theo}{Theorem}

\newtheorem{Lem}{Lemma}
\newtheorem{Coro}{Corollary}

\def\traitfinal{\nobreak\bigskip\nobreak%
\hbox to\textwidth{\hfil\hbox to 8 cm{\hrulefill}\hfil}}

\def\Abs#1{\left\vert #1 \right\vert}   

\def\CV_#1^#2{\mathrel{
\mathop{\kern 0pt\hbox to 10mm{\rightarrowfill}}
\limits_{#1}^{#2}}}

\def\E{\mathop{\hbox{E}}\nolimits}              
\def\Var{\mathop{\hbox{Var}}\nolimits}          

\def\Eqalign#1{\null\,\vcenter{\openup\jot\m@th\ialign{
\strut\hfil$\displaystyle{##}$&$\displaystyle{{}##}$\hfil
&&\quad\strut\hfil$\displaystyle{##}$&$\displaystyle{{##}}$
\hfil\crcr#1\crcr}}\,}

\def\system#1{\left\{\null\,\vcenter{\openup1\jot\m@th%
\ialign{\strut\hfil$##$&$##$\hfil&&\enspace$##$\enspace&%
\hfil$##$&$##$\hfil\crcr#1\crcr}}\right.}

\newcommand{\ldeux}{$L^2([0,1])$}
\newcommand{\somi}{\sum_{i\leq h_n}}
\newcommand{\somr}{\sum_{r=1}^{k_n}}
\newcommand{\aikn}{a_{i,k_n}}
\newcommand{\xnret}{X_{n,r}^{\star}}
\newcommand{\znret}{Z_{n,r}^{\star}}

\newcommand{\ynr}{Y_{n,r}}
\newcommand{\unr}{u_{n,r}}

\newcommand{\inr}{I_{n,r}}
\newcommand{\anr}{\alpha_{n,r}}
\newcommand{\tnr}{t_{n,r}}
\newcommand{\phinr}{\phi_{n,r}}
\newcommand{\Fnr}{F_{n,r}}
\newcommand{\Fnrb}{\bar{F}_{n,r}}

\newcommand{\Dnr}{D_{n,r}}

\newcommand{\Mnr}{M_{n,r}}
\newcommand{\mnr}{m_{n,r}}
\newcommand{\Bu}{B_{n,1}(x)}
\newcommand{\Bd}{B_{n,2}(x)}
\newcommand{\Bdd}{B_{n,2}^2(x)}
\newcommand{\Bdt}{B_{n,2}^3(x)}

\newcommand{\Btt}{B_{n,3}^3(x)}
\newcommand{\Bi}{B_{n,\infty}(x)}

\newcommand{\lnr}{\lambda_{n,r}}

\newcommand{\aiknchap}{\hat a_{i,k_n}}
\newcommand{\fnchap}{\hat f_n}
\newcommand{\fnchapx}{\hat f_n(x)}
\newcommand{\gnchapx}{\hat g_n(x)}

\newcommand{\proof}{{\bf Proof~: }}
\newcommand{\CQFD}
{%
\mbox{}%
\nolinebreak%
\hfill%
\rule{2mm}{2mm}%
\medbreak%
\par%
}
\newcommand{\normsup}[1]
{{\parallel{#1}\parallel}_{\infty}} 
 
\newcommand{\normdeux}[1]
{{\parallel{#1}\parallel}_2} 
\newcommand{\grando}[1]
{O\left({#1}\right)} 
\newcommand{\petito}[1]
{o\left({#1}\right)} 
\newcommand{\indic}[1] 
{{\bf 1}_{#1}}

\begin{document}

\title{Projection estimates of point processes boundaries}
\author{St\'ephane Girard \& Pierre Jacob\\\\
Laboratoire de Probabilit\'es et Statistique, Universit\'e Montpellier 2,\\
place Eug\`ene Bataillon, 34095 Montpellier cedex 5, France.\\
{\tt \{girard, jacob\}@stat.math.univ-montp2.fr}}

\date{}

\maketitle

\begin{abstract}
We present a method for estimating the edge of a two-dimensional
bounded set, given a finite random set of points drawn from the interior.
The estimator is based both on projections on $C^1$ bases and on extreme points of the point process.
We give conditions on the Dirichlet's kernel associated to the $C^1$ bases
for various kinds of convergence and asymptotic normality.
We propose a method for reducing the negative bias and illustrate it by a simulation.
\end{abstract}

\section{Introduction}

We address the problem of estimating a bounded set $S$ of ${\mathbf R}^2$ given
a finite random set $\Sigma$ of points drawn from the interior. This kind of problem arises in various frameworks such as 
classification~\cite{HarRas}, image processing~\cite{KorTsy2} or econometrics
problems~\cite{DepSimTul}. A lot of different solutions were proposed since
\cite{Geff1} and \cite{RenSul} depending on the properties of the observed random set $\Sigma$
and of the unknown set $S$.
In this paper, we focus on the special case where $\Sigma$ 
is the set of points of an homogeneous Poisson
process whose support is $S=\{(x,y)\in {\bf R}^2\mid 0\leq x\leq 1~;\;0\leq y \leq f(x)\}$,
where $f$ is an unknown function. Thus, the estimation of subset $S$
reduces to the estimation of function $f$. Let us note that this kind of
support was already considered in \cite{Geff1}.
In the wide range of nonparametric functional estimators \cite{Bosq}, piecewise
polynomials have been especially studied \cite{KorTsy,KorTsy2,MamTsy,Gay,Har}
and their asymptotic optimality was established under different regularity conditions on $f$.\\
The first support estimator based on orthogonal series appears in \cite{AbbSuq}. Its properties
are extensively studied in \cite{JacSuq} in the case of Haar and $C^1$ bases.
The expansion coefficients estimation requires the knowledge of the process
intensity. This is a limitation which is avoided in \cite{GirJac} in the case of the Haar basis
by considering a coefficient estimation based on the extreme points of the sample.
In this paper, a similar study is carried out in the case of $C^1$ bases.
The estimator can be written as a linear combination of extreme values involving the 
Dirichlet's kernel of the $C^1$ basis. A close study of the extreme values 
stochastic properties as well as a precise control of the Dirichlet's kernel
behavior allow one to establish general conditions for various convergences and
asymptotic normality of the estimator. Our results are illustrated for the
trigonometric basis case. 
Note that the model proposed here is well-adapted to the estimation of 
a bounded star-shaped subset of the plane.
Let $D$ be such a domain. Then, there
exists a convex subset $C$ of this domain, called the kernel of $D$, from
which the whole boundary $\partial D$ of $D$ can be seen.
If we assume an interior point of $C$ to be known, the use of polar coordinates
allows to reduce the problem of estimating $\partial D$ to the problem
considered here, that is the estimation of a function $f$, with the particularity that $f(0)=f(1)$.
In such a situation, extreme points observed in the neighborhood
of $x=0$ bring information on the behavior of $f$ in the neighborhood
of $x=1$ and {\it vice versa}. Thus, an estimation relying on the
trigonometric basis is specially well adapted.\\
This paper is organized as follows. Section \ref{defs} is devoted to
the definition of the estimator and Section \ref{basic} presents some
basic results on extreme values and Dirichlet's kernels.
The mean integrated square convergence of the estimate
is briefly studied in Section \ref{sectionestim} and the
asymptotic normality is established in Section \ref{sectionnor}.
In Section \ref{sectionreduc} a very simple bias correction is proposed,
illustrated in~\cite{Nous} by a simulation.

\section{Definition of the estimator}
\label{defs}

\subsection{Preliminaries}

Let $N$ be a Poisson process with a mean measure $\mu=c\lambda$,
where the intensity parameter $c$ is unknown, $\lambda$ is the Lebesgue measure, and the support of $N$ is given by:
\begin{equation}
\label{defS}
 S=\{(x,y)\in {\bf R}^2\mid 0\leq x\leq 1~;\;0\leq y \leq f(x)\}.
\end{equation}
We assume that $f$ is measurable and satisfies
\begin{equation}\label{mM}
0<m=\inf_{[0,1]}f \leq M=\sup_{[0,1]}f<+\infty,
\end{equation}
which entails the square integrability of $f$ on $[0,1]$.
In the sequel, we will introduce extra hypothesis on $f$ as needed.\\
 Let $(e_i)_{i\in {\bf N}}$ be an orthonormal basis of {\ldeux.} 
The expansion of $f$ with respect to the basis 
is supposed to be both $L^2$ and pointwise convergent to $f$ on
$[0,1]$:   
\begin{equation}
\label{serie1}
\forall x \in [0,1],\;\; f(x)= \sum_{i=0}^{+\infty}a_ie_i(x), 
\end{equation}
with
\begin{equation}
\label{serie2}
\forall i \geq 0,\;\; a_i=\int_0^1 e_i(t)f(t)\,dt.
\end{equation}
We denote by $K_n$ the
Dirichlet's kernel associated to the orthonormal basis $(e_i)_{i\in {\bf N}}$ defined by
\begin{equation}
K_n(x,y)=\sum_{i=0}^{h_n} e_i(x)e_i(y), \qquad (x,y)\in [0,1]^2, 
\end{equation}
where $(h_n)$ is an increasing sequence of integers.
The trigonometric basis will provide us with an important example
to illustrate our convergence results. It is defined by
\begin{equation}
e_0(x)=1,\qquad e_{2k-1}(x)=\sqrt{2}\cos{2k\pi x},\qquad
e_{2k}(x)=\sqrt{2}\sin{2k\pi x},
\quad k\geq 1,
\end{equation}
and we shall suppose for convenience that $h_n$ is even. This leads to 
\begin{equation}
\label{defnoyau}
 K_n(x, y)=
\left|
\begin{array}{ll}
\displaystyle\frac{\sin{(1+h_n)\pi (x-y)}} {\sin{\pi (x-y)}} & x\neq y,\\
1+h_n & x=y.\label{diagtrigo}
\end{array}
\right.
\end{equation}

\noindent The speed at which the sequence $(a_k)$ decreases to 0 is linked to the regularity of $f$. 
In the case of the trigonometric basis, if $f$ is a function of class $C^2$ then
\begin{equation}
\label{eqai}
 a_k=\grando{k^{-2}},
\end{equation}
(see \cite{Gasquet}).
The estimator is built in two steps. First, in subsection \ref{step1}, $f$ is
approximated by a sequence $(f_n)$ obtained from its expansion with respect to
the orthogonal
basis. Then, in subsection \ref{step2}, an estimator $\fnchap$ of $f_n$ is
proposed.

\subsection{Approximation of $f$}
\label{step1}

Let $(k_n)$ be an increasing sequence of non-negative integers such that 
$k_n=\petito{n}$. Divide $S$ into $k_n$ cells $D_{n,r}$ where:
\begin{equation}
D_{n,r}= \{\;(x,y) \in S \mid x\in I_{n,r}\;\}, \qquad
I_{n,r}=\left[\frac{r-1}{k_n}, \frac{r}{k_n}\right[,\quad r=1,\ldots,k_n.
\end{equation} 
Each coefficient $a_i$ is approximated by discretizing (\ref{serie2})
according to:
\begin{equation}
\aikn = \somr  e_i(x_r)\lambda (D_{n,r}), \qquad x_r=\frac{2r-1}{2k_n}.
\end{equation}
Then, the expansion (\ref{serie1}) is truncated to the $h_n$ first terms leading to 
\begin{equation}
f_n(x)=\somi \aikn e_i(x), \qquad x\in [0,1],
\end{equation}
which can be written in terms of the Dirichlet's kernel:
\begin{equation}
f_n(x)=\somr K_n(x_r,x) \lambda(D_{n,r}), \qquad x\in [0,1].
\end{equation}
Let us emphasize that the approximation $f_n$ of $f$ only depends on the basis 
$(e_i)_{i\in {\bf N}}$ through its Dirichlet's kernel. 
The next step towards the definition of the estimator consists of
estimating $\lambda (D_{n,r})$.

\subsection{Estimation of $f_n$}
\label{step2}

Let $N^{*n}$ denote the superposition $N_1+\cdots +N_n $ of $n$ independent
copies of the point process $N$, and $\Sigma_n$ the set of points generated
by $N^{*n}$. For $r=1,\dots,k_n$, consider 
the maximum $\xnret$ of the second coordinates of the set of points
$\Sigma_{n,r}=\Sigma_n \cap \Dnr$. Of course, if $\Sigma_{n,r}=\emptyset$,
set $\xnret=0$. 
Then, $\lambda(D_{n,r})$ can be estimated by $\xnret/k_n$. 
This leads to an estimate $\aiknchap$ of $\aikn$ defined as:
\begin{equation}
\label{defaiknchap}
\aiknchap =\somr e_i(x_r)\frac{\xnret}{k_n}, \qquad 1\leq i \leq h_n, 
\end{equation}
and consequently to an estimate $\fnchapx$ of $f_n(x)$ via:
\begin{equation}
\fnchapx  = \somi \aiknchap e_i(x)= \somr K_n(x_r,x) \frac{\xnret}{k_n}.
\end{equation}
Two remarks can be made. First, the estimator does not require knowledge
of $c$, which ensures a wide range of applications. Second, the estimator
is written as a linear combination of the maxima $\xnret$ involving
the Dirichlet's kernel. Thus, the analysis of the behavior of $\fnchap$
will rely on both studies of the Dirichlet's kernel features
and of the maxima's stochastic properties. This is the topic of the
next section.

\section{Basic results}
\label{basic}

\subsection{Bounds on the Dirichlet's kernel}

For $x\in [0,1]$ and $j\in\{1,2,3\}$ define
$$
B_{n,j}(x)=\left(\somr \Abs{K_n(x_r,x)}^j\right)^{1/j}
$$
and
$$
B_{n,\infty}(x)=\max_{1\leq r\leq k_n} \Abs{K_n(x_r,x)}.
$$
In what follows,  some theorems involving conditions
on $B_{n,j}(x)$ for $j\in\{1,2,3,\infty\}$ are fiven. Some
of these conditions follow easily from the properties of $K_n$,
see~\cite{JacSuq}. In the following lemma two
properties which are less straightforward are given for the special case
of the trigonometric basis.
\begin{Lem}
\label{lemtri}
Suppose $K_n$ is the Dirichlet's kernel associated to the trigonometric basis.
\begin{enumerate}
\item [(i)] If $h_n=\petito{k_n}$ then $\displaystyle \sup_x \Bu = \grando{k_n\ln h_n}$,
\item [(ii)] If $h_n\ln h_n=\petito{k_n}$, then for all $x\in[0,1]$, $\Bd \sim (k_n h_n)^{1/2}$.
\end{enumerate}
\end{Lem}
The proof is postponed to the Appendix.

\subsection{Maxima stochastic properties}

In the sequel, we write:
$$
\lambda (D_{n,r})=\lnr,\; \min_{x\in\inr} f(x) = \mnr,\; \max_{x\in\inr} f(x) = \Mnr.
$$
Recall that $\xnret$ is the maximum of the second coordinates of the set of points
$\Sigma_{n,r}$. \\
Noticing that, for $0\leq x\leq\mnr$,
\begin{equation}
P(\xnret\leq x)=P(N^{\star n}(\Dnr\setminus(\inr\times[0,x]))=0),
\end{equation}
we easily obtain 
the distribution function $\Fnr(x)=P(\xnret\leq x)$ on $[0,\mnr]$:
\begin{equation}
\label{eqfdr}
\Fnr(x)= \exp{\left[ \frac{nc}{k_n} (x-k_n\lnr)\right]}.
\end{equation}
Straight forward calculations lead to the following expansions for the
mathematical expectation and the variance of $\xnret$, where 
the knowledge of the $C^1$-regularity of $f$ compensates for the
lack of a precise expression for $\Fnr$ on $]\mnr,\Mnr[$.
\begin{Lem}
\label{lemespe} 
Suppose $f$ is a function of class $C^1$, $n=\petito{k_n^2}$ and
$k_n=\petito{{n}/{\ln n}}$. Then,
\begin{enumerate}
\item [(i)] $\E(\xnret)= k_n\lnr -\displaystyle \frac{k_n}{nc} + \petito{\frac{n}{k_n^3}}$,
\item [(ii)] $\Var(\xnret)\sim \displaystyle \frac{k_n^2}{n^2 c^2}$.
\end{enumerate}
\end{Lem}
The proof of the following lemma, which is more difficult, is postponed
to the appendix.
\begin{Lem} \label{lemphi}
Suppose $f$ is a function of class $C^1$ and $k_n=\petito{{n}/{\ln n}}$.
Let $(\tnr)$ be a sequence such that $\tnr=\petito{{n}/{k_n}}$ and
$\tnr=\petito{{k_n^3}/{n}}$.
Then, the characteristic function of $(\xnret-\mnr)$ can be written at point $\tnr$ as:
$$
\phinr(\tnr)= \frac{1 + i\tnr \frac{k_n}{nc} \Fnrb(\mnr) + 
\petito{\Abs{\tnr}\frac{n}{k_n^3}} + \petito{n^{-s}}}{1+ i\tnr \frac{k_n}{nc}},
$$
with $s$ arbitrary large, and $\Fnrb=1-\Fnr$.
\end{Lem}

\section{Estimate convergences }
\label{sectionestim}

We refer to \cite{JacSuq} for a careful study of the bias convergences
$\normdeux{f_n-f}\to 0$ and $\normsup{f_n-f}\to~0$.
Then, we just have to consider $(\fnchap-f_n)$. A complete investigation
for mean integrated convergence, mean uniform convergence,
$L^2$-almost complete convergence and uniform almost complete convergence
is available at our website. In order to avoid lengthy developments,
we just propose the following basic result.
\begin{Theo}
Suppose $f$ is a $C^1$ function.
If $\normdeux{f_n-f}=\petito{1}$ and if 
$\displaystyle \sup_x \Bu=\petito{n^2/k_n}$,
then $\E(\normdeux{\fnchap-f})=\petito{1}$.
\end{Theo}
\proof
Introducing the random variable $\ynr=({\xnret}/{k_n})-\lnr$, we have
\begin{eqnarray}
\E\left(\normdeux {\fnchap - f_n}^2\right) &=&
 \E\left( \sum_{r,s} Y_{n,r} K_n(x_r,x_s) Y_{n,s}\right), \\
&\leq& 2 \sum_{r,s} \Abs{K_n(x_r,x_s)} ( \E(Y_{n,r}^2) + \E(Y_{n,s}^2) )\\
&\leq& 4 \sup_x \Bu \somr \E(Y_{n,r}^2),
\end{eqnarray}   
and the result follows from Lemma \ref{lemespe}.
\CQFD
\begin{Coro}
If $K_n$ is the Dirichlet's kernel associated to the trigonometric basis
and $f$ is a $C^1$ function then
$h_n\ln^{1/2} h_n =\petito{k_n}$ and $k_n (\ln h_n)^{1/2}=\petito{n}$
 are sufficient conditions for
$\E(\normdeux{\fnchap-f})=\petito{1}$.
\end{Coro}
\proof
From \cite{JacSuq}, Proposition 3, $h_n (\ln h_n)^{1/2}=\petito{k_n}$
entails $\normdeux{f_n-f}=\petito{1}$.
Moreover, from Lemma~\ref{lemtri}(i), if $k_n (\ln h_n)^{1/2}=\petito{n}$,
then
\begin{equation}
\sup_n \Bu =\grando{k_n\ln h_n}=\petito{n^2/k_n},
\end{equation}
and the conclusion follows.
\CQFD

\section{Asymptotic distribution}
\label{sectionnor}

In this section, we present a limit theorem for the distribution
of $(\fnchap-\E\fnchap)$. A similar result is not available
for $(\fnchap-f)$ without reducing the bias, which is done in the
next section.
\begin{Theo}
\label{thnorasymp}
Suppose that $f$ is a function of class $C^1$. 
If  $k_n=\petito{{n}/{\ln n}}$, $n=\petito{k_n^{3/2}}$ and
$\Bi=\petito{\Bd}$,
 then
$(nc/\Bd)(\fnchapx - \E(\fnchapx))$
converges in distribution to a standard Gaussian variable for all $x\in[0,1]$.
\end{Theo}
\proof
Denote $\anr=\mnr-{k_n}/{(nc)}$ and for all $x\in[0,1]$ introduce
$\psi_{n,x}$ the characteristic function of $(nc/\Bd)(\fnchapx - \E(\fnchapx))$.  It expands as:
\begin{eqnarray}
\label{decomp}
\psi_{n,x}(t) & = & \exp{it\left(\frac{nc}{k_n\Bd}\somr K_n(x_r,x) (\anr-\E(\xnret)) \right)} \\
\label{decomp2}
&\times& \E\left[\exp{it\left(\frac{nc}{k_n\Bd}\somr K_n(x_r,x)(\xnret - \anr) \right)}\right].
\end{eqnarray}
Consider first the argument of (\ref{decomp}):
\begin{equation}
T_n(x)=\frac{nc}{k_n\Bd}\somr K_n(x_r,x) (\anr-\E(\xnret)),
\end{equation}
and show it converges to 0. By the Cauchy-Schwartz inequality:
\begin{equation}
\Abs{T_n(x)}  \leq  \frac{nc}{k_n} \frac{\Bu}{\Bd} \max_r \Abs{\anr-\E(\xnret)}
	\leq	 \frac{nc}{k_n^{1/2}} \max_r \Abs{\anr-\E(\xnret)}.
\end{equation}
Now Lemma \ref{lemespe} entails
\begin{equation}
\Abs{T_n(x)} \leq  \frac{n}{k_n^{1/2}}\left(\max_r (k_n\lnr-\mnr)+\petito{\frac{n}{k_n^3}}\right)
= \petito{\frac{n}{k_n^{3/2}}},
\end{equation}
which converges to 0.
Thus, introducing $\tnr=\frac{nc  K_n(x_r,x) t}{k_n\Bd}$, we obtain:
\begin{eqnarray}
\psi_{n,x}(t)&\!\!\!\! \sim\!\!\!\! &\exp{it\left(\frac{1}{\Bd} \somr K_n(x_r,x) \right)}
\E\left[\exp{it\left(\frac{nc}{k_n\Bd}\somr K_n(x_r,x)(\xnret-\mnr)\right)}\right]
\nonumber \\
&\!\!\!\!=\!\!\!\!& \exp{it\left(\frac{1}{\Bd} \somr K_n(x_r,x) \right)}
\prod_{r=1}^{k_n}\phinr(\tnr).
\end{eqnarray}
To apply Lemma \ref{lemphi}, we have to verify that $\tnr=\petito{{n}/{k_n}}$
and $\tnr=\petito{{k_n^3}/{n}}$.
The first condition is satisfied since
\begin{equation}
\label{maj1}
\Abs{\tnr\frac{k_n}{nc}}\leq \Abs{t}\frac{\Bi}{\Bd}=\petito{1}.
\end{equation}
The second condition is satisfied as well:
\begin{equation}
\label{maj2}
\Abs{\tnr\frac{n}{k_n^3}} = \Abs{\tnr\frac{k_n}{n}} \frac{n^2}{k_n^4}=\petito{1}.
\end{equation}
The characteristic function can be seen to be of the order:
\begin{eqnarray}
\label{part3}
\psi_{n,x}(t) & \!\!\! \sim\!\!\!  & \frac{\exp{it\displaystyle \left(\frac{1}{\Bd} \somr K_n(x_r,x) \right)}}
{ \displaystyle \prod_{r=1}^{k_n} \left(1+i\tnr\frac{k_n}{nc}\right)} \\
& \!\!\! \times\!\!\!  &
\displaystyle\prod_{r=1}^{k_n}\left( 1 + i\tnr \frac{k_n}{nc}\Fnrb(\mnr) +
\petito{\Abs{\tnr}\frac{n}{k_n^3}} + \petito{n^{-s}} \right).
\label{part4}
\end{eqnarray}
Consider first the logarithm of term (\ref{part3}). A second-order Taylor expansion yields
\begin{equation}
J_n^{(1)}(x)=\somr\left[ i\tnr \frac{k_n}{nc}-\ln{\left( 1 + i\tnr \frac{k_n}{nc}\right)}\right] 
=-\frac{t^2}{2} + \grando{\frac{\Btt}{\Bdt}}.
\end{equation}
Since ${\Btt}/{\Bdt}\leq {\Bi}/{\Bd}=\petito{1}$, 
it follows that $J_n^{(1)}(x)\to -t^2/2$ as $n\to\infty$.\\
Finally, consider the logarithm of (\ref{part4}): 
\begin{equation}
J_n^{(2)}=\somr \ln{(1+\unr)}, \mbox{ with } \unr=  i\tnr \frac{k_n}{nc}\Fnrb(\mnr) +\petito{\Abs{\tnr}\frac{n}{k_n^3}} + \petito{n^{-s}}.
\end{equation}
Observe that $\max_r \Abs{\unr}$ converges to 0 from (\ref{maj1}) and (\ref{maj2}).
Thus, for $n$ large enough $\Abs{\unr}~<~1/2$ uniformly in $r$ and the
classical identity $\Abs{\ln(1+\unr)}<\Abs{\unr}$ yields
\begin{equation}
\Abs{J_n^{(2)}}\leq \Abs{t}\frac{\Bu}{\Bd}\frac{nc}{k_n^{2}} + \petito{\frac{\Bu}{\Bd}}\frac{n^2}{k_n^{4}} + \petito{k_n n^{-s}}.
\end{equation}
Therefore, the Cauchy-Schwartz inequality leads to $J_n^{(2)} \to 0$ and  $\psi_{n,x}(t) \to e^{-t^2/2}$ as $n\to\infty$.
\CQFD

\begin{Coro}
\label{cornorasymp}
Suppose $K_n$ is the Dirichlet's kernel associated to the trigonometric basis
and $f$ is a $C^1$ function. 
If $h_n =\petito{k_n}$, $k_n=\petito{{n}/{\ln n}}$ and $n=\petito{k_n^{3/2}}$, then, for all $x\in[0,1]$,
$nc(h_n k_n)^{-1/2}(\fnchapx - \E(\fnchapx))$
converges in distribution to a standard Gaussian variable. 
\end{Coro}
\proof
From (\ref{defnoyau}), $\Bi\leq \normsup{K_n}=1+h_n$, and from
Lemma \ref{lemtri}(ii), $\Bd\sim (h_nk_n)^{1/2}$ so that
$
{\Bi}/{\Bd}=\petito{(h_n/k_n)^{1/2}}=\petito{1}.
$
\CQFD
\noindent 
Possible choices of $k_n$ and $h_n$ sequences in Corollary \ref{cornorasymp}
are $k_n=n^{2/3}(\ln n)^\varepsilon$
and $h_n=(\ln n)^\varepsilon$ for $\varepsilon>0$ arbitrary small.
These choices entail $n(h_n k_n)^{-1/2}=n^{2/3}(\ln n)^{-\varepsilon}$.

\section{Reducing the bias}
\label{sectionreduc}

The bias can be decomposed as follows:
\begin{equation}
\E(\fnchap-f)=(\E\fnchap-f_n) + (f_n-f),
\end{equation}
where the first term in the sum is the statistical part of the bias,
and the second term is the systematic part of the bias.
First, consider the irreductible bias $(f_n-f)$. In order to obtain
a limit distribution for $(nc/B_{n,2}(x))(\fnchapx-f(x))$, we need
to satisfy the condition
\begin{equation}
\label{condition}
\lim_{n\to\infty} \frac{nc}{B_{n,2}(x)}(f_n(x)-f(x)) = 0.
\end{equation}
Introduce $\displaystyle S_n(f)=\sum_{i=0}^{h_n} a_i e_i$.
Equation (3.12) in \cite{JacSuq} provides sharp bounds for 
$(S_n(f)-f_n)$, so that the question reduces to considering
$(S_n(f)-f)$, which only depends on the basis.
We shall see that the trigonometric basis satisfies
(\ref{condition}) under reasonable conditions on $h_n$ and $k_n$.
In the case of a general $C^1$ basis, we shall take (\ref{condition})
as a condition.\\
Now, it follows from Lemma \ref{lemespe}(i) that
\begin{equation}
\E\fnchapx-f_n(x) = \somr K_n(x_r,x)\left( \frac{\E(\xnret)}{k_n}-\lnr \right)
\end{equation}
presents a negative component which should be eliminated.
To this end, for $r=1,\dots,k_n$, define $\znret$ by
$\znret=0$ if $\Sigma_{n,r}=\emptyset$ and 
$\znret$ is the infimum of the second coordinates of the
points of $\Sigma_{n,r}$ otherwise.
Then, the random variable
\begin{equation}
\label{defzn}
Z_n=\frac{1}{k_n}\somr \znret,
\end{equation}
has a mathematical expectation
\begin{equation}
\label{espeZ}
\E(Z_n)=\frac{k_n}{nc} + \petito{\frac{n}{k_n^3}},
\end{equation}
and a variance 
\begin{equation}
\Var(Z_n)\sim\frac{k_n}{n^2c^2},
\end{equation}
(use Lemma \ref{lemespe} for $\znret$).
We define a corrected estimate by
\begin{equation}
\tilde{f}_n(x)=\sum_{r=1}^{k_n} K_n(x_r,x)\left(\frac{\xnret+Z_n}{k_n}\right)
=\fnchapx + \gnchapx.
\end{equation}
\begin{Lem}
\label{lemsup2}
Suppose $f$ is a function of class $C^1$, $n=\petito{k_n^2}$ and
$k_n=\petito{{n}/{\ln n}}$. Then,
$$
\frac{nc}{\Bd}\Abs{\E(\tilde{f}_n(x))-f_n(x)}=\petito{n^2/k_n^{7/2}},\;
\forall x\in[0,1].
$$
If, moreover, for all $x\in[0,1]$, $k_n=\petito{\Bdd}$ and (\ref{condition}) holds,
then
$$
\frac{nc}{\Bd}(\hat{g}_n(x)-\E(\hat{g}_n(x)))=o_P(1).
$$
\end{Lem}
\proof
We have 
\begin{eqnarray}
\frac{nc}{\Bd}\Abs{\E(\tilde{f}_n(x))-f_n(x)}
&\leq & \frac{nc}{k_n}\frac{\Bu}{\Bd}
\max_{1\leq r\leq k_n} \Abs{\E(\xnret)+\E(Z_n)-k_n\lnr} \\
&=& \frac{nc}{k_n}\frac{\Bu}{\Bd} \grando{\frac{n}{k_n^3}}
= \grando{\frac{n^2}{k_n^{7/4}}},
\end{eqnarray}
from (\ref{espeZ}), Lemma \ref{lemespe} and the Cauchy-Schwartz inequality.\\
Now, applying (\ref{condition}) to the constant function $f=1$ yields
\begin{equation}
\frac{1}{k_n}\somr K_n(x_r,x) \to 1,
\end{equation}
as $n\to\infty$. Therefore, $\Var(\hat{g}_n(x))\sim \Var(Z_n)\sim k_n/(n^2c^2)$
 and then
\begin{equation}
\Var\left(\frac{nc}{\Bd}(\hat{g}_n(x)-\E(\hat{g}_n(x))\right)\sim\frac{k_n}{\Bdd} 
\end{equation}
which converges to 0.
\CQFD
\begin{Theo}
\label{thnorasymp2}
Suppose that $f$ is a function of class $C^1$. If the following
conditions are verified
\begin{enumerate}
\item [(i)] $k_n=\petito{{n}/{\ln n}}$, $n=\petito{k_n^{3/2}}$,
\item [(ii)] for all $x\in[0,1]$, $\max (k_n^{1/2}, \Bi) = \petito{\Bd}$,
\item [(iii)] for all $x\in[0,1]$, $\displaystyle \frac{nc}{\Bd}\Abs{f_n(x)-f(x)}=\petito{1}$,
\end{enumerate}
 then, for all $x\in[0,1]$,
$(nc/\Bd)(\tilde{f}_n(x) - f(x))$
converges in distribution to a standard Gaussian variable.
\end{Theo}
The proof is a simple consequence of the expansion
\begin{equation}
(\tilde{f}_n-f)=(\fnchap-\E(\fnchap))+(\hat{g}_n-\E(\hat{g}_n))
+(E(\tilde{f}_n)-f_n)+(f_n-f),
\end{equation}
and of Theorem \ref{thnorasymp} and Lemma \ref{lemsup2}. 
\begin{Coro}
\label{cornorasymp2}
Suppose $K_n$ is the Dirichlet's kernel associated to the trigonometric basis
and $f$ is a $C^2$ function.
If $h_n\ln{h_n}=\petito{k_n}$, $n=\petito{h_n^{3/2}k_n^{1/2}}$, $nh_n^{1/2}\ln{h_n}=\petito{k_n^{3/2}}$  and $k_n=\petito{{n}/{\ln n}}$  then,
for all $x\in[0,1]$, $nc(h_nk_n)^{-1/2}(\tilde{f}_n(x) - f(x))$
converges in distribution to a standard Gaussian variable.
\end{Coro}
\proof
Conditions $(i)$, $(ii)$ of Theorem \ref{thnorasymp2} are verified.
Consider $(iii)$. On using (\ref{eqai}) we have
\begin{eqnarray}
 nc (h_n k_n)^{-1/2} \Abs{S_n(f)(x)-f(x)}
& \leq & nc (h_n k_n)^{-1/2} \sum_{i\geq h_n} |a_i| |e_i(x)|  \nonumber \\
& \leq & \sqrt{2} nc (h_n k_n)^{-1/2} \sum_{i\geq h_n} |a_i|   \nonumber \\
& \leq & \sqrt{2} nc (h_n k_n)^{-1/2} \sum_{i\geq h_n} i^{-2}.
\end{eqnarray}
A straightforward calculation yields
\begin{equation}
\label{fin1}
\frac{nc}{\Bd} \Abs{S_n(f)(x)-f(x)} =\grando{n h_n^{-3/2} k_n^{-1/2}}.
\end{equation}
From \cite{JacSuq}, equations (3.11) and (3.12),
\begin{equation}
\Abs{S_n(f)(y)-f_n(y)}=\grando{\frac{1}{k_n}\int_0^1 \Abs{\frac{\partial K_n}{\partial x}(v,y)}dv}
\end{equation}
and
\begin{equation}
\int_0^1 \Abs{\frac{\partial K_n}{\partial x}(v,y)}dv = \grando{h_n\ln h_n}.
\end{equation}
Therefore,
\begin{equation}
\label{fin2}
\frac{nc}{B_{n,2}(y)} \Abs{S_n(f)(y)-f_n(y)} = \grando{n k_n^{-3/2} h_n^{1/2} \ln h_n }
 \end{equation}
and (\ref{fin1}) with (\ref{fin2}) conclude the proof.
\CQFD

\noindent 
Possible choices of $k_n$ and $h_n$ sequences in Corollary \ref{cornorasymp2}
are $k_n=n^{4/5}(\ln n)^{3/5}(\ln\ln n)^\varepsilon$
and $h_n=n^{2/5}(\ln n)^{-1/5}(\ln\ln n)^\varepsilon$ for $\varepsilon>0$ arbitrary small.
These choices entail $n(h_n k_n)^{-1/2}=n^{2/5}(\ln n)^{-1/5}(\ln\ln n)^{-\varepsilon}$.

\section{Conclusion and further developments}

In this paper, we showed how the convergence results established in
\cite{GirJac} in the case of the Haar basis can be adapted for any
$C^1$ basis under some assumptions on the Dirichlet's kernel behavior.
 We have emphasized that the estimator and these assumptions
only depend on the Dirichlet's kernel of the basis. This suggests
to define a new estimator based on a Parzen-Rosenblatt kernel.

\newpage
\section*{Appendix}

\subsection*{Proof of Lemma \ref{lemtri}}
\begin{enumerate}
\item [(i)] In the sequel, we shall use the inequality found in \cite{JacSuq},
equation~(2.11):
\begin{equation}
\label{majosin}
\Abs{\frac{\sin(pu)}{\sin(u)}} \leq p \indic{[0,\delta]}(\Abs{u})
+ \frac{\pi}{2\Abs{u}} \indic{[\delta,\pi/2]}(\Abs{u}),
\end{equation}
for all $p>0$, $0< \delta < \pi/2$ and $\Abs{u}<\pi/2$. 
Taking account of the periodicity and symmetry properties of the
trigonometric kernel, it suffices to study
\begin{equation}
\sup_{x\in [0,1/k_n]} \frac{2}{k_n} \sum_{r=1}^{[k_n/2]+1} \Abs{K_n(x_r,x)},
\end{equation}
where $[u]$ denotes the integer part of $u$.
Let us write
\begin{equation}
\frac{1}{k_n} \sum_{r=1}^{[k_n/2]+1} K_n(x_r,x) =
\frac{1}{k_n} \sum_{r=1}^{[\gamma_n]} K_n(x_r,x) +
\frac{1}{k_n} \sum_{r=[\gamma_n]+1}^{[k_n/2]+1} K_n(x_r,x),
\end{equation}
with $\gamma_n=k_n/(h_n+1)$, and consider the two terms separately.
\begin{itemize}
\item Introduce $\delta=(\pi/k_n)([\gamma_n]-1/2)$.
For $r=1,\dots,[\gamma_n]$, we have $\pi(x_r-x)\leq \delta$ and thus
(\ref{majosin}) yields $\Abs{K_n(x_r,x)}\leq 1+h_n$ which  
gives in turn
\begin{equation}
\label{parti}
\frac{1}{k_n} \sum_{r=1}^{[\gamma_n]} \Abs{K_n(x_r,x)}\leq 1. 
\end{equation}
\item For $r=[\gamma_n]+1,\dots,[k_n/2]+1$, we have
$\pi(x_r-x)\geq \delta$ and consequently
(\ref{majosin}) yields
\begin{equation}
\frac{1}{k_n} \sum_{r=[\gamma_n]+1}^{[k_n/2]+1} \Abs{K_n(x_r,x)}
\leq \frac{1}{k_n} \sum_{r=[\gamma_n]+1}^{[k_n/2]+1} \frac{1}{2(x_r-x)}
\leq \frac{1}{2}\frac{1}{k_n} \sum_{r=[\gamma_n]}^{[k_n/2]}
\frac{1}{\frac{1}{k_n}\left(r-\frac{1}{2}\right)}.
\end{equation}
Therefore, 
\begin{equation}
\label{partii}
\frac{1}{k_n} \sum_{r=[\gamma_n]+1}^{[k_n/2]+1} \Abs{K_n(x_r,x)}
\leq \frac{1}{2} \int_{\frac{\delta}{\pi}}^{\frac{1}{2}+\frac{1}{2k_n}}
\frac{du}{u} \leq \frac{1}{2}\ln(4(h_n+1)),
\end{equation}
for $k_n>2(h_n+1)$.
\end{itemize}
Finally, collecting (\ref{parti}) and (\ref{partii}), we obtain
\begin{equation}
\sup_{x\in [0,1]} \frac{1}{k_n} \sum_{r=1}^{k_n} \Abs{K_n(x_r,x)}
\leq 2 +  \ln(4(h_n+1)).
\end{equation}
\item [(ii)] 
From \cite{JacSuq}, equation (4.14), 
\begin{equation}
\label{ajout1}
\Abs{\frac{\Bdd}{k_n K_n(x,x)}- 1} \leq \frac{\normsup{K_n}}{K_n(x,x)}
\frac{1}{k_n} \sup_x \int_0^1 \Abs{\frac{\partial K_n}{\partial y}(x,v)}dv.
\end{equation}
In the case of the trigonometric basis $K_n(x,x)=\normsup{K_n}=1+h_n$
(see (\ref{defnoyau})) and from \cite{JacSuq}, equation (3.12) we have
\begin{equation}
\label{ajout2}
\int_0^1 \Abs{\frac{\partial K_n}{\partial y}(x,v)}dv = \grando{h_n\ln h_n}=\petito{k_n}.
\end{equation}
The conclusion follows from (\ref{ajout1}) and (\ref{ajout2}).
\end{enumerate}
\CQFD

\subsection*{Proof of Lemma \ref{lemphi}}
Consider the expansion
\begin{equation}
\E(e^{i\tnr\xnret})=P(\xnret=0)+\int_0^{\mnr}e^{ix\tnr}\Fnr'(x)dx + \int_{\mnr}^{\Mnr}
 e^{ix\tnr} \Fnr(dx).
\end{equation}
The first and second term can be computed explicitely since (\ref{eqfdr}) provides an expression of $\Fnr$ on $[0,\mnr]$:
\begin{equation}
\label{tmp1}
\E(e^{i\tnr\xnret})=e^{-nc\lnr} + \frac{ e^{i\tnr\mnr} \Fnr(\mnr) - e^{-nc\lnr}} 
{1 + i\tnr\frac{k_n}{nc}} + \int_{\mnr}^{\Mnr} e^{ix\tnr} \Fnr(dx).
\end{equation}
The third term can be expanded as
\begin{equation}
\label{tmp2}
\int_{\mnr}^{\Mnr} e^{ix\tnr} \Fnr(dx)= e^{i\tnr\mnr}\Fnrb(\mnr)
+\int_{\mnr}^{\Mnr} (e^{ix\tnr}-e^{i\tnr\mnr})\Fnr(dx),
\end{equation}
with $\Abs{e^{ix\tnr}-e^{i\tnr\mnr}}\leq (\Mnr-\mnr)\Abs{\tnr}$. Thus
\begin{eqnarray}
\Abs{\int_{\mnr}^{\Mnr} (e^{ix\tnr}-e^{i\tnr\mnr})\Fnr(dx)} 
& \leq & (\Mnr-\mnr)\Abs{\tnr}\Fnrb(\mnr) \nonumber \\
& \leq & (\Mnr-\mnr)\Abs{\tnr} \frac{nc}{k_n}(k_n\lnr-\mnr)\nonumber \\
\label{tmp3}
& = & \grando{\Abs{\tnr}\frac{n}{k_n^3}}.
\end{eqnarray}
Collecting (\ref{tmp1})--(\ref{tmp3}), 
and remarking that $k_n=\petito{{n}/{\ln n}}$
yields $\Abs{\frac{k_n}{nc} e^{-nc\lnr}}=\grando{n^{-s}}$ concludes the proof. 
\CQFD


\newpage

\begin{thebibliography}{xx}


\bibitem{AbbSuq}{ Abbar, H. and Suquet, Ch. } (1993)
{ Estimation $L^2$ du contour d'un processus de Poisson homog\`ene sur le plan. }
{\em Pub. IRMA Lille}, {\bf 31}, {\bf II}.

\bibitem{Bosq} { Bosq, D.} (1977)
{ Contribution \`a la th\'eorie de l'estimation fonctionnelle.}
{\em Publications de l'Institut de Statistique de l'Universit\'e de Paris},
{\bf XIX}, 2, 1--96.





\bibitem{DepSimTul}{ Deprins, D., Simar, L. and Tulkens, H.} (1984)
{ Measuring Labor Efficiency in Post Offices.} in
{\em The Performance of Public Enterprises: Concepts and Measurements} by
M. Marchand, P. Pestieau and H. Tulkens, North Holland ed, Amsterdam.

\bibitem{Gasquet} { Gasquet, C., Witomski, P.} (1990)
{\em Analyse de Fourier et Applications}, Masson.

\bibitem{Gay} { Gayraud, G.} (1997)
{ Estimation of functionals of density support.}
{\em Math. Methods Statist.}  {\bf 6}(1), 26--46.

\bibitem{Geff1} { Geffroy, J.} (1964)
{ Sur un probl\`eme d'estimation g\'eom\'etrique.}
{\em Publications de l'Institut de Statistique de l'Universit\'e de Paris},
 {\bf XIII}, 191--200.


\bibitem {GirJac}Girard, S. and Jacob, P. (2003)   Extreme
values and Haar series estimates of point processes boundaries.
\textit{Scandinavian Journal of Statistics}, \textbf{30}(2), 369--384.

\bibitem{Nous} { Girard, S. and Jacob, P.} (2003)
Projection estimates of point processes boundaries.
\textit{Journal of Statistical Planning and Inference}, \textbf{116}(1), 1--15.


\bibitem{Har}{ H\"ardle, W., Park, B. U. and Tsybakov, A. B.} (1995)
{ Estimation of a non sharp support boundaries.}
{\em J. Multiv. Analysis}, {\bf 43}, 205--218.

\bibitem{HarRas}{ Hardy A. and Rasson J. P.} (1982)
{ Une nouvelle approche des probl\`emes de classification automatique.}
{\em Statistique et Analyse des donn\'ees}, {\bf 7}, 41--56.


\bibitem{JacAbb} { Jacob, P. and Abbar, H.} (1989)
{ Estimating the edge of a Cox process area.}
{\em Cahiers du Centre d'Etudes de Recherche Op\'erationnelle}, {\bf 31}, 215--226.

\bibitem{JacSuq} { Jacob, P. and Suquet, P.} (1995)
{ Estimating the edge of a Poisson process by orthogonal series.}
{\em Journal of Statistical Planning and Inference}, {\bf 46},
 215--234.


\bibitem{KorTsy}{ Korostelev, A., Simar, L. and Tsybakov, A. B.} (1995)
{ Efficient estimation of monotone boundaries.}
{\em The Annals of Statistics}, {\bf 23}, 476--489.

\bibitem{KorTsy2}{  Korostelev, A. P. and Tsybakov, A. B.} (1993)
{ Minimax linewise algorithm for image reconstruction.}
in {\em Computer intensive methods in statistics} by W. H\"ardle, L. Simar ed.,
Statistics and Computing, Physica Verlag, Springer.

\bibitem{MamTsy}{ Mammen, E. and Tsybakov, A. B.} (1995)
{ Asymptotical minimax recovery of set with smooth boundaries.}
{\em The Annals of Statistics}, {\bf 23}(2), 502--524.

\bibitem{RenSul}{ Renyi, A. and Sulanke, R.} (1963)
{  Uber die konvexe H\"ulle von n zuf\"alligen gew\"ahlten Punkten.}
{\em Z. Wahrscheinlichkeitstheorie verw. Geb.} {\bf 2}, 75--84.




\end{thebibliography}
\end{document}